\documentstyle[twoside,epsfig]{article}

\catcode`\@=11
\long\def\@makefntext#1{
\protect\noindent \hbox to 3.2pt {\hskip-.9pt  
$^{{\eightrm\@thefnmark}}$\hfil}#1\hfill}		

\def\@makefnmark{\hbox to 0pt{$^{\@thefnmark}$\hss}}	
	
\def\ps@myheadings{\let\@mkboth\@gobbletwo
\def\@oddhead{\hbox{}
\rightmark\hfil\eightrm\thepage}   
\def\@oddfoot{}\def\@evenhead{\eightrm\thepage\hfil
\leftmark\hbox{}}\def\@evenfoot{}
\def\sectionmark##1{}\def\subsectionmark##1{}}

\oddsidemargin=\evensidemargin
\newcounter{sectionc}\newcounter{subsectionc}\newcounter{subsubsectionc}
\renewcommand{\section}[1] {\vspace{12pt}\addtocounter{sectionc}{1} 
\setcounter{subsectionc}{0}\setcounter{subsubsectionc}{0}\noindent 
	{\tenbf\thesectionc. #1}\par\vspace{5pt}}
\renewcommand{\subsection}[1] {\vspace{12pt}\addtocounter{subsectionc}{1} 
	\setcounter{subsubsectionc}{0}\noindent 
	{\bf\thesectionc.\thesubsectionc. {\kern1pt \bfit #1}}\par\vspace{5pt}}
\renewcommand{\subsubsection}[1] {\vspace{12pt}\addtocounter{subsubsectionc}{1}
	\noindent{\tenrm\thesectionc.\thesubsectionc.\thesubsubsectionc.
	{\kern1pt \tenit #1}}\par\vspace{5pt}}

\newcounter{appendixc}
\newcounter{subappendixc}[appendixc]
\newcounter{subsubappendixc}[subappendixc]
\renewcommand{\thesubappendixc}{\Alph{appendixc}.\arabic{subappendixc}}
\renewcommand{\thesubsubappendixc}
	{\Alph{appendixc}.\arabic{subappendixc}.\arabic{subsubappendixc}}

\renewcommand{\appendix}[1] {\vspace{12pt}
        \refstepcounter{appendixc}
        \setcounter{figure}{0}
        \setcounter{table}{0}
        \setcounter{lemma}{0}
        \setcounter{theorem}{0}
        \setcounter{corollary}{0}
        \setcounter{definition}{0}
        \setcounter{equation}{0}
        \renewcommand{\thefigure}{\Alph{appendixc}.\arabic{figure}}
        \renewcommand{\thetable}{\Alph{appendixc}.\arabic{table}}
        \renewcommand{\theappendixc}{\Alph{appendixc}}
        \renewcommand{\thelemma}{\Alph{appendixc}.\arabic{lemma}}
        \renewcommand{\thetheorem}{\Alph{appendixc}.\arabic{theorem}}
        \renewcommand{\thedefinition}{\Alph{appendixc}.\arabic{definition}}
        \renewcommand{\thecorollary}{\Alph{appendixc}.\arabic{corollary}}
        \renewcommand{\theequation}{\Alph{appendixc}.\arabic{equation}}
        \noindent{\tenbf Appendix \theappendixc #1}\par\vspace{5pt}}
\newcommand{\subappendix}[1] {\vspace{12pt}
        \refstepcounter{subappendixc}
        \noindent{\bf Appendix \thesubappendixc. {\kern1pt \bfit #1}}
	\par\vspace{5pt}}
\newcommand{\subsubappendix}[1] {\vspace{12pt}
        \refstepcounter{subsubappendixc}
        \noindent{\rm Appendix \thesubsubappendixc. {\kern1pt \tenit #1}}
	\par\vspace{5pt}}

\newcommand{\textlineskip}{\baselineskip=13pt}
\newcommand{\smalllineskip}{\baselineskip=10pt}

\def\eightcirc{
\begin{picture}(0,0)
\put(4.4,1.8){\circle{6.5}}
\end{picture}}
\def\eightcopyright{\eightcirc\kern2.7pt\hbox{\eightrm c}} 

\newcommand{\copyrightheading}[1]
	{\vspace*{-2.5cm}\smalllineskip{\flushleft
	{\footnotesize International Journal of Modern Physics A, #1}\\
	{\footnotesize $\eightcopyright$\, World Scientific Publishing
	 Company}\\
	 }}

\def\abstracts#1#2#3{{
	\centering{\begin{minipage}{4.5in}\baselineskip=10pt\footnotesize
	\parindent=0pt #1\par 
	\parindent=15pt #2\par
	\parindent=15pt #3
	\end{minipage}}\par}} 

\renewenvironment{thebibliography}[1]
	{\frenchspacing
	 \ninerm\baselineskip=11pt
	 \begin{list}{\arabic{enumi}.}
	{\usecounter{enumi}\setlength{\parsep}{0pt}
	 \setlength{\leftmargin 12.7pt}{\rightmargin 0pt} 
	 \setlength{\itemsep}{0pt} \settowidth
	{\labelwidth}{#1.}\sloppy}}{\end{list}}

\newcounter{itemlistc}
\newcounter{romanlistc}
\newcounter{alphlistc}
\newcounter{arabiclistc}

\newcommand{\fcaption}[1]{
        \refstepcounter{figure}
        \setbox\@tempboxa = \hbox{\footnotesize Fig.~\thefigure. #1}
        \ifdim \wd\@tempboxa > 5in
           {\begin{center}
        \parbox{5in}{\footnotesize\smalllineskip Fig.~\thefigure. #1}
            \end{center}}
        \else
             {\begin{center}
             {\footnotesize Fig.~\thefigure. #1}
              \end{center}}
        \fi}

\newcommand{\tcaption}[1]{
        \refstepcounter{table}
        \setbox\@tempboxa = \hbox{\footnotesize Table~\thetable. #1}
        \ifdim \wd\@tempboxa > 5in
           {\begin{center}
        \parbox{5in}{\footnotesize\smalllineskip Table~\thetable. #1}
            \end{center}}
        \else
             {\begin{center}
             {\footnotesize Table~\thetable. #1}
              \end{center}}
        \fi}

\def\@citex[#1]#2{\if@filesw\immediate\write\@auxout
	{\string\citation{#2}}\fi
\def\@citea{}\@cite{\@for\@citeb:=#2\do
	{\@citea\def\@citea{,}\@ifundefined
	{b@\@citeb}{{\bf ?}\@warning
	{Citation `\@citeb' on page \thepage \space undefined}}
	{\csname b@\@citeb\endcsname}}}{#1}}

\newif\if@cghi
\def\cite{\@cghitrue\@ifnextchar [{\@tempswatrue
	\@citex}{\@tempswafalse\@citex[]}}
\def\citelow{\@cghifalse\@ifnextchar [{\@tempswatrue
	\@citex}{\@tempswafalse\@citex[]}}
\def\@cite#1#2{{$\null^{#1}$\if@tempswa\typeout
	{IJCGA warning: optional citation argument 
	ignored: `#2'} \fi}}

\def\pmb#1{\setbox0=\hbox{#1}
	\kern-.025em\copy0\kern-\wd0
	\kern.05em\copy0\kern-\wd0
	\kern-.025em\raise.0433em\box0}

\def\fnt#1#2{\footnotetext{\kern-.3em
	{$^{\mbox{\scriptsize #1}}$}{#2}}}

\def\fpage#1{\begingroup
\voffset=.3in
\thispagestyle{empty}\begin{table}[b]\centerline{\footnotesize #1}
	\end{table}\endgroup}

\def\runninghead#1#2{\pagestyle{myheadings}
\markboth{{\protect\footnotesize\it{\quad #1}}\hfill}
{\hfill{\protect\footnotesize\it{#2\quad}}}}
\font\tenrm=cmr10
\font\tenit=cmti10 
\font\tenbf=cmbx10
\font\bfit=cmbxti10 at 10pt
\font\ninerm=cmr9

\font\eightrm=cmr8

\def\qed{\hbox{${\vcenter{\vbox{			
   \hrule height 0.4pt\hbox{\vrule width 0.4pt height 6pt
   \kern5pt\vrule width 0.4pt}\hrule height 0.4pt}}}$}}

\begin{document}
\addtolength{\textheight}{.5cm}

\runninghead{Top Production and Decay at Linear Colliders
 $\ldots$} {Top Production and Decay at Linear Colliders $\ldots$}

\normalsize\textlineskip
\thispagestyle{empty}
\setcounter{page}{1}

\copyrightheading{}			

\vspace*{0.88truein}
\vspace*{-9\baselineskip}
\begin{flushright}
\hfill {\rm UR-1620}\\
\hfill {\rm ER/40685/958}\\
\hfill {\rm December 2000}\\
\end{flushright}
\vspace*{2\baselineskip}

\fpage{1}
\centerline{\bf BFKL PHYSICS IN JET PRODUCTION AT  $e^+e^-$ 
COLLIDERS\footnote{Presented at the 2000 
Meeting of the Division of Particles and Fields of the APS, Columbus,
OH, August 9--12, 2000.}}
\vspace*{0.37truein}
\centerline{\footnotesize LYNNE H. ORR}
\vspace*{0.015truein}
\centerline{\footnotesize\it Department of Physics and Astronomy, 
University of Rochester}
\baselineskip=10pt
\centerline{\footnotesize\it Rochester NY 14627-0171, USA}
\vspace*{10pt}
\centerline{\footnotesize W.J. STIRLING}
\vspace*{0.015truein}
\centerline{\footnotesize\it Departments of Physics and Mathematical 
Sciences and IPPP} 
\baselineskip=10pt
\centerline{\footnotesize\it University of Durham, Durham DH1 3LE, UK}

\vspace*{0.21truein}
\abstracts{Virtual photon scattering in $e^+e^-$ collisions can 
result in events with the electron-positron pair at large rapidity
separation with hadronic activity in between. The BFKL equation resums 
large logarithms that dominate the cross
section for this process. We report here on a Monte Carlo method 
for solving the BFKL equation that allows
kinematic constraints to be taken into account and show results 
for $e^+e^-$ collisions. }{}{}


\vspace*{1pt}\textlineskip	
\section{Leading-Order BFKL and Improved BFKL Monte Carlo} 
\vspace*{-0.5pt}
\noindent
In the QCD description of hadronic processes there can
appear logarithms that multiply each power of the strong coupling 
constant $\alpha_s$, spoiling the utility of fixed order perturbation
theory in the regions where these logarithms are large.
The BFKL equation\cite{bfkl}  resums 
large logarithms  due to emission of multiple soft
gluons (real and virtual)  which are comparable in transverse momentum but 
strongly ordered in rapidity.  

The BFKL equation can be solved analytically if there are no constraints
(e.g.\ from kinematics) on the transverse momenta of 
emitted gluons.  Clearly this requires giving up conservation of energy,
and furthermore the implicit sum over arbitrary numbers of 
gluons gives a result with leading-order kinematics only.  
For
purposes of comparison with experiment, a more realistic BFKL
prediction would be better.  This can be obtained\cite{os,schmidt} by solving
the BFKL equation iteratively, making the gluon summation explicit.
The result can be incorporated into a Monte Carlo event generator,
allowing for kinematic constraints to be applied directly.  
The Monte Carlo approach has been applied to dijet production
at large rapidity in hadron colliders and it improves the agreement 
between BFKL predictions and experiment\cite{os}.

\section{Virtual Photon Scattering at $e^+e^-$ Colliders}

BFKL physics can be important in virtual photon scattering into 
hadrons at $e^+e^-$ colliders when the electron and positron are
scattered into the forward and backward regions (``double-tagged'' events).
BFKL effects are relevant in the region where the invariant mass $W$ 
of the hadronic system is large and $$s>>Q^2>>\Lambda_{QCD}^2,$$
where $Q^2$ is the photon virtuality and $\sqrt{s}$ is the total
c.m.\ energy.  The corresponding fixed-order QCD process
is $\gamma^*\gamma^*\to q\bar{q}q\bar{q}$ via $t$-channel gluon
exchange.  The BFKL result is obtained by attaching a gluon ladder
to the $t$-channel gluon.
\begin{figure}
\epsfxsize200pt
\psfig{figure=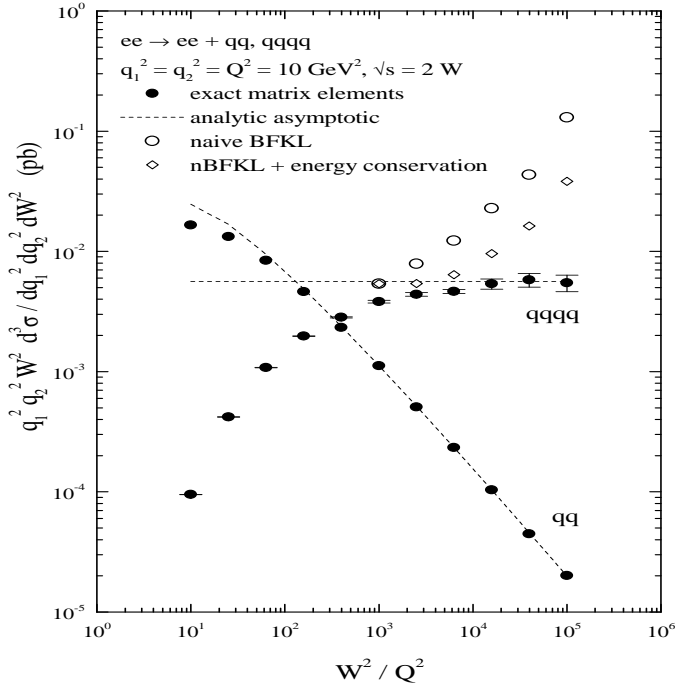,width=10cm,height=10cm}
\caption{Exact (closed data points) and analytic asymptotic (dashed line) 
$e^+e^- \to e^+e^-q \bar q$  and 
$e^+e^- \to e^+e^-q \bar q q \bar q$  cross sections versus 
$W^2/Q^2$ at fixed $W^2/s = 1/4$.  Also shown:  analytic BFKL without (open 
circles) and with (open diamonds) energy conservation imposed.} 
\end{figure}

Figure 1 compares the fixed-order QCD cross section with the naive 
(analytic) BFKL prediction.  The cross section shown,
$$
W^2 Q_1^2 Q_2^2 \; {d^3 \sigma \over  d W^2 d Q_1^2 d Q_2^2 } 
$$
as a function of $W^2/Q^2$ for fixed $\sqrt{s}/W$  approaches a 
constant in fixed-order QCD but the BFKL cross section (open
circles) continues to rise.  The improved BFKL Monte Carlo calculation
will result in a slower rise; that calculation is in progress \cite{osprog}but 
an upper limit can be obtained by imposing an upper limit on the 
gluon energies, as shown with open diamonds.  It should be 
noted that the position in $W^2/Q^2$ where BFKL is matched
to asymptotic QCD is arbitrary in leading order; our choice is reasonable 
but not fixed in the theory.

\begin{figure}
\vspace*{-3cm}
\psfig{figure=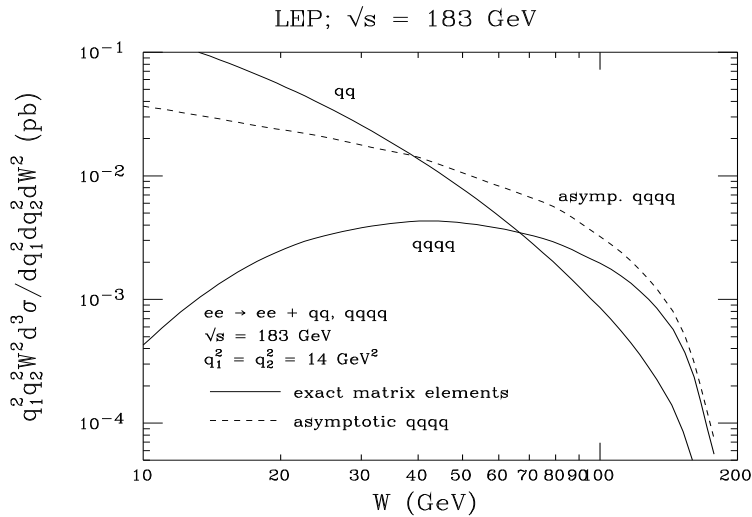,width=11cm,height=14cm}
\vskip -3cm
\caption{Exact (solid lines) and analytic asymptotic (dashed line) 
$e^+e^- \to e^+e^-q \bar q$  and 
$e^+e^- \to e^+e^-q \bar q q \bar q$  cross sections versus 
$W^2/Q^2$ at fixed $\sqrt{s}=183\ {\rm GeV}$.} 
\end{figure}

The L3 collaboration at LEP have measured the $\gamma^*\gamma^*$ cross
section from couble-tagged $e^+e^-$ events, and their
result lies between asymptotic QCD (which is flat) and analytic
BFKL (which rises); see for example\cite{LEP}.  It is likely
that the BFKL Monte Carlo prediction will lie closer to the 
data, but the QCD prediction itself deserves some scrutiny.  
Figure 2 compares the exact and asymptotic QCD predictions for 
the LEP energy $\sqrt{s}=183\ {\rm GeV}$  (this is 
the undivided $e^+e^-$ cross section that includes the photon luminosity
and is therefore not flat).  The LEP measurements correspond to 
values of $W$ in the range between about 15 and 90 GeV.  We see from the 
figure that the QCD prediction is not close to its asymptotic limit,
and the ratio of the two rises in this region, as do the data.
Until 
the fixed-order QCD and BFKL Monte Carlo predictions are sorted out,
it is not clear what we can  conclude from the data.
This work, at both LEP and  
linear collider energies, is currently in progress\cite{osprog}.

\section{Acknowledgments}
\noindent
Work supported in part by the U.S. Department of Energy and the U.S. 
National Science Foundation,
under grants DE-FG02-91ER40685 and  PHY-9600155.

\end{document}